\documentclass[reprint, aip, jcp,
floatfix]{revtex4-2}

\usepackage[T1]{fontenc}  
\usepackage[utf8]{inputenc}  
\usepackage{amsmath, amssymb}  
\usepackage{mathcmd}  
\usepackage[version=4]{mhchem}  
\usepackage[dvipsnames]{xcolor}  
\usepackage{xspace}  
\usepackage{soul}  
\usepackage{hyperref}  
\hypersetup{  
    colorlinks,  
    linkcolor={RoyalBlue},  
    citecolor={RoyalBlue},  
    urlcolor={RoyalBlue}  
}  
\usepackage{graphicx}  
\usepackage{setspace}  
\usepackage{tikz}  

\newcommand{\dgwol}{\Delta G_{\textnormal{W}\rightarrow\textnormal{Ol}}}

\newcommand{\pkatext}{p$K_{\textnormal{a}}$\xspace}

\newcommand{\apkatext}{ap$K_{\textnormal{a}}$\xspace}  
\newcommand{\apkamath}{\textnormal{ap}K_{\textnormal{a}}}  

\newcommand{\bpkatext}{bp$K_{\textnormal{a}}$\xspace}  
\newcommand{\bpkamath}{\textnormal{bp}K_{\textnormal{a}}}  

\newcommand{\phtext}{pH\xspace}  
\newcommand{\phmath}{\textnormal{pH}}  

\newcommand{\kbt}{k_\textnormal{B}T}

\newcommand{\etal}{\textit{et al}.\xspace}  
\newcommand{\ie}{i.e.,\xspace}

\begin{document}

\title{Data-driven equation for drug--membrane permeability across drugs and membranes}

\author{Arghya Dutta}
\email{dutta@mpip-mainz.mpg.de}
\affiliation{Max Planck Institute for Polymer Research, Mainz, Germany}

\author{Jilles Vreeken}
\affiliation{CISPA Helmholtz Center for Information Security, Saarbr\"ucken, Germany}

\author{Luca M. Ghiringhelli}
\affiliation{The NOMAD Laboratory at the Fritz Haber Institute of the Max Planck
Society and Humboldt University, Berlin, Germany}

\author{Tristan Bereau}
\affiliation{Van 't Hoff Institute for Molecular Sciences and Informatics
	Institute, University of Amsterdam, Amsterdam, The Netherlands}
\affiliation{Max Planck Institute for Polymer Research, Mainz, Germany}

\date{\today}

\begin{abstract}
	Drug efficacy depends on its capacity to permeate across the cell membrane.
	We consider the prediction  of passive drug--membrane permeability
	coefficients. Beyond the widely recognized correlation with hydrophobicity,
	we additionally consider the functional relationship between passive
	permeation and acidity. To discover easily interpretable equations that
	explain the data well, we use the recently proposed sure-independence
	screening and sparsifying operator (SISSO), an artificial-intelligence
	technique that combines symbolic regression with compressed sensing. Our
	study is based on a large in silico dataset of 0.4 million small molecules
	extracted from coarse-grained simulations. We rationalize the equation
	suggested by SISSO via an analysis of the inhomogeneous
	solubility--diffusion model in several asymptotic acidity regimes. We
	further extend our analysis to the dependence on lipid-membrane composition.
	Lipid-tail unsaturation plays a key role, but surprisingly contributes
	stepwise rather than proportionally. Our results are in line with previously
	observed changes in permeability, suggesting the distinction between
	liquid-disordered (Ld) and liquid-ordered (Lo) permeation. Together,
	compressed sensing with analytically derived asymptotes establish and
	validate an accurate, broadly applicable, and interpretable equation for
	passive permeability across both drug and lipid-tail chemistry.
\end{abstract}
\maketitle

\section{Introduction}
Passive lipid-membrane permeation is and remains of great relevance for
pharmaceutical applications and an improved physicochemical understanding of
small-sized molecules in complex biological materials.\cite{Lipinski2001,
avdeef2012absorption} The technological implications of the problem has
sustained the need for experiment- and simulation-free prediction of passive
permeation that are rapid, inexpensive, and accurate.\cite{di2015drug,
Broccatelli2016} Various types of surrogate models have been proposed over the
years, the field having adopted machine learning early on.\cite{Bodor1991} While
modern deep-learning approaches take advantage of unchallenged model
expressivity to offer unprecedented prediction accuracy, they suffer from two
important drawbacks:
\begin{enumerate}
	\item Overfitting: The size of chemical space of drug-like molecules is
	      overwhelmingly large ($\sim 10^{60}$ compounds).\cite{Dobson2004} Deep
	      surrogate models need large numbers of parameters to establish complex
	      relationships. Unfortunately the body of experimentally available data
	      is minuscule compared to the size of chemical space. This can lead to
	      surrogate models that shift dangerously upon addition/removal of small
	      numbers of compounds in the training set. The problem is aggravated by
	      databases that are often proprietary, preventing broad availability
	      and reproducibility. Relying on different measurement batches tends to
	      also accentuate systematic errors.
	\item Lack of interpretability: Surrogate models are oftentimes black-box
	      techniques that typically cloud \emph{why} a certain prediction has
	      been made. Deep neural networks exhibit an overwhelming number of
	      parameters and rely on highly non-linear hierarchical functions,
	      making them nearly impossible to conceptually grasp.\cite{Gilpin2018}
	      Quantitative structure--activity relationship (QSAR) models can build
	      multivariate models, but the combination of too many descriptors will
	      lead to similar difficulties. Beyond predicting individual data
	      points, we seek to gain further insight. Insight is essential, for
	      instance as a stepping stone to solving the inverse problem, thereby
	      establishing structure--property relationships and enabling compound
	      design.
\end{enumerate}
In this work, we address these points using a combination of approaches.
Critically, we address overfitting by relying on large datasets applied to
simple models. Instead of experimental approaches, we base our study on \emph{in
silico} measurements, taking advantage of the rapid rise of high-throughput
molecular dynamics simulations.\cite{Bereau2020} All-atom simulations offer a
gold standard in terms of simulation-based permeability modeling that can reach
exquisite correlation against experimental reference, but their overwhelming
computational costs unfortunately limit studies to tens of
compounds.\cite{orsi2010passive, carpenter2014method, lee2016simulation} Here
instead, we use an approach based on particle-based coarse-grained (CG)
simulations, making use of the transferable Martini
model.\cite{marrink2004coarse, marrink2007martini, marrink2013perspective,
alessandri2020martini}  The Martini model sacrifices some chemical resolution,
but retains the essential driving force, mainly the partitioning coefficient of
a chemical group between different phases. This allows the CG approach to
recover excellent accuracy: 1.4~kcal/mol along a potential of mean force (PMF),
translating to $1~\log_{10}$ unit in the permeability coefficient, validated
across an extensive set of structurally distinct compounds against both
atomistic simulations and experimental measurements.\cite{menichetti2019drug}
Critically, the accuracy of the CG model is accompanied by a
three-orders-of-magnitude speedup compared to atomistic simulations. The
high-throughput coarse-grained (HTCG) simulations offer unprecedentedly large
databases of permeability coefficients: Menichetti \etal reported results for
$511\,427$ compounds.\cite{menichetti2019drug} The ability to screen over so
many compounds results mainly from the transferable nature of the coarse-grained
model---many molecules map to the same set of CG beads, which effectively
\emph{reduces} the size of chemical space.\cite{Kanekal2019} The benefits for an
efficient computational-screening procedure outweigh the impinged degeneracy, as
indicated by the above-mentioned accuracy. Overall, the database contains a
nearly exhaustive subset of small organic molecules in the range 30--160\,Da,
thereby ensuring a dense coverage of the chemical space in this subset. A deeper
analysis of this large database is the subject of this study.

As for the data-driven model, we explicitly avoid building and using a black-box
model, and instead turn to learning an \emph{equation}. In particular, we will
rely on recently proposed data-driven techniques to discover
equations.\cite{Ghiringhelli2015, Brunton2016} Equations relevant to physical
problems often display simplifying properties, such as symmetries or
separability easing both their data-driven discovery and generalization beyond
the training set. Several studies have demonstrated the ability to (re)discover
physics equations.\cite{schmidt2009distilling, Udrescu2020} Generalization
capabilities are critical, because typical training datasets are minuscule
relative to the size of chemical compound space, such that overfitting can
easily prevail. How do we discover simple equations from the data? To this end,
we follow Occam's razor and limit the complexity of the equations we consider.
The combination of descriptors through various mathematical operators will lead
to an overwhelming number of trial equations, many of which may fit the data
similarly well. Following ideas from compressed sensing, we select \emph{simple}
equations by applying the $\ell_0$ regularization. Identifying simple equations
both benefits generalization aspects, but also interpretability, \ie
rationalizing the derived equation.

Passive drug permeation measures the propensity of a solute to spontaneously
cross a lipid membrane. In this paper, we exclude transporter-mediated uptake to
focus solely on thermal diffusion.\cite{Kell2014} Upon doing so, the solute
interacts with a great variety of physicochemical environments---from an
aqueous phase to a hydrophobic membrane core. Permeation is quantified by means
of its coefficient, $P$, as the steady state flux of the solute across the soft
interface. Early on, Meyer\cite{meyer1899theorie} and
Overton\cite{overton1901studien} modeled passive permeation as diffusion across
a homogeneous slab via $P = KD/\sigma$, where $K$ and $D$ are the water/membrane
partitioning coefficient and diffusivity of the compound, respectively, and
$\sigma$ is the thickness of the bilayer core. $K$ is typically approximated by
a simpler proxy, namely the partitioning coefficient between water and octanol.
Water/octanol partition, or more generally hydrophobicity, has long been
identified as strongly correlating with membrane permeability.\cite{Levin1980}
Notable refinements to the homogeneous Meyer--Overton rule include the
inhomogeneous solubility--diffusion model (ISDM), estimating the permeability
coefficient via an integral over the membrane extension, $z$, of its PMF,
$G(z)$, as
\begin{equation}
	\label{eq:isdm}
	P^{-1} = \int {\rm d}z\, R(z) = \int {\rm d}z\,\frac{\exp[\beta G(z)]}{D(z)},
\end{equation}
where $R(z)$ and $\beta = 1/\kbt$ correspond to an associated resistivity and
the inverse temperature, respectively.\cite{Diamond1974} The competition between
different protonation states naturally follows the sum of inverse resistivities,
analogous to the total resistance in a parallel electrical
circuit.\cite{carpenter2014method} PMFs are shifted according to the difference
between the solution's \phtext and the compound's acid dissociation constant,
\pkatext. In the following, we take the perspective of a neutral compound, which
can deprotonate (acidic, \apkatext) or protonate (basic, \bpkatext). Knowledge
of the PMF(s) and the diffusivity thereby fully determines the permeability
coefficient. Unfortunately, these ($i$) are so far only available by
computational techniques and ($ii$) typically require extensive calculations
($\sim 10^5$\,CPU-hour per system at an atomistic
resolution).\cite{orsi2010passive, lee2016simulation}

Even at a CG resolution, Eq.~\ref{eq:isdm} still requires free-energy
calculations to determine the PMF. The objective of this study is to gain
further insight into the key physical determinants of the permeability
coefficient. Beyond the widely known effect of hydrophobicity, we focus on
incorporating the role of acidity via the compound's protonation state.
\cite{Radak2017} The role of acidity is crucial, partly because of the sheer
number of ionizable drugs: they make $\approx 62$\% of the World Health
Organization's list of essential drugs. \cite{Manallack2007, Manallack2013} It
has long been hypothesized that the neutral form of an ionizable drug is the
only contributor to its permeability, known as the pH--partition hypothesis.
\cite{Shore1957, Saparov2006} However, this hypothesis is limiting for two
reasons. First, permanently charged compounds are known to permeate lipid
membranes. \cite{Langguth1997, Fischer2007} Second, differences between a
compound's \pkatext and the surrounding pH can lead to a protonation-coupled
permeation mechanism, which calls for the combined contributions of the neutral
and ionized forms. \cite{Palm1999, Yue2019} Clearly the role of acidity is
expected to couple with hydrophobicity. Establishing a \emph{functional}
relationship connecting the two quantities is the objective of this work. To
this end, we rely on modern data-driven techniques to discover an equation.
Relating acidity to the permeability coefficient would not only help
establishing rapid estimates for ionic compounds, but also offer insight into
the coupling of these physicochemical properties that are valid for a wide class
of compounds.

Limitations in the number of candidate descriptors and correlations between
features have recently been addressed by sure-independence screening and
sparsifying operator (SISSO), \cite{ouyang2018sisso, ouyang2019simultaneous}
which is an artificial-intelligence technique that combines symbolic regression
with compressed sensing.\cite{donoho2006compressed, candes2006robust,
baraniuk2007compressive} SISSO provides a data-driven framework to discover
equations---mathematical relations between input variables that best correlate
with the target property. We will discuss several equations of various
complexities to illustrate the balance between accuracy and interpretability. We
root the simplest variant in the underlying physics by a comparison against
analytical acid--base asymptotic regimes. This simple model incorporating both
hydrophobicity and acidity allows us to easily extend our analysis to different
lipid membranes starting from limited information. From knowledge of neutral
species alone, we predict the change in passive permeability in various lipid
membranes. We finally discuss the change in permeability in the context of
membrane-phase behavior.

\section{Methods}

\subsection{Database of drug--membrane thermodynamics}
Our analysis is based on the passive-permeability database provided by
Menichetti \etal\cite{menichetti2019drug} Reference information includes
water/octanol partitioning free energy ($\dgwol$), acid dissociation constant
for acids and bases (\apkatext and \bpkatext), and simulation-based permeability
coefficient (expressed by its order of magnitude, $\log_{10}P$) for solutes
through a single-component bilayer made of
1,2-dioleoyl-\textit{sn}-glycero-3-phosphocholine (DOPC). The water/octanol
partitioning free energies were predicted using the neural network
ALOGPS,\cite{tetko2002application} and acid dissociation constants \pkatext were
predicted from ChemAxon Marvin.\cite{marvin} Filtering of the Generated DataBase
(GDB)\cite{fink2005virtual} for compounds that mapped to a one- or two-bead
coarse-grained Martini representation, \ie~a monomer or a dimer, led to
$511\,427$ small organic molecules ($30$--$160$~Da). Enhanced-sampling molecular
dynamics simulations yielded the PMFs for both neutral and (de)protonated
species, and Eq.~\ref{eq:isdm} was used to compute the permeability coefficient,
$P$. The diffusivity profile was \emph{not} extracted from coarse-grained
simulations, but instead from previous atomistic studies, taking advantage of
its relatively small chemical dependence and its logarithmic impact on the
permeability.\cite{carpenter2014method} The \pkatext of a chemical group can be
either acidic (\apkatext) or basic (\bpkatext), in that a neutral compound can
either deprotonate or protonate (see SI for definitions). While the ionization
constant of conjugated acid/base pairs typically lie between $10^{-2}$ and
$10^{16}$, we considered compounds with \pkatext values between $-10$ and
$20$.\cite{dixon1993estimation} This led to a dataset of $418\,828$ compounds
used as part of this work. A follow up work to Menichetti \etal provided PMFs
for the same set of neutral Martini small molecules in different phosphocholine
(PC) lipid membranes: 1,2-diarachidonoyl-\emph{sn}-glycero-3-PC (DAPC);
1,2-dilinoleoyl-\emph{sn}-glycero-3-PC (DIPC);
1,2-dilauroyl-\emph{sn}-glycero-3-PC (DLPC); 1,2-dioleoyl-\emph{sn}-glycero-3-PC
(DOPC); 1,2-dipalmitoyl-\emph{sn}-glycero-3-PC (DPPC); and
1-palmitoyl-2-oleoyl-\emph{sn}-glycero-3-PC (POPC).\cite{Hoffmann2020}

\subsection{Learning algorithm}
To learn an interpretable model of passive permeability, we used SISSO, as
implemented in Ref.~\citenum{ouyang2017sisso}. SISSO aims at establishing a
functional relationship, $y=f(\Phi)$, between $n$ primary features,
$\Phi_0=\{\phi_1, \phi_2, \cdots, \phi_n\}$, and a target property, $y$, based
on $N$ training compounds. SISSO assumes that $y$ can be reliably expressed as a
linear combination of non-linear, but closed-form, functions of primary
features. To construct these non-linear functions, SISSO recursively applies a
set of user-defined unary and binary operators (we used $\{+, \, -, \, \times,
\, \div, \, \exp, \, \log, \, \;()^{-1}, \,\;()^{2}, \,\;()^{3}, \,
\sqrt[3]{()}, \, \sqrt{()}\}$) on the primary features and builds up sets of
candidate features. $\Phi_q$ denotes the set of candidate features at each level
of recursion $q$. The number of candidate features in $\Phi_q$ increases sharply
with increase in the recursion level $q$, the number of operators used, and the
number of primary features $n$. For each $q$, SISSO selects iteratively subsets
of candidate features that have the largest linear correlations with the target
$y$ and then with the subsequent residuals, \ie each portion of $y$ that is
captured by the previous iterations (see Ouyang
\etal\cite{ouyang2019simultaneous}). The number of iterations in this procedure,
which equals the number of terms in the linear expansion of $f(\Phi)$, and
hereby denoted as dimension of the model, is controlled by a sparsifying
$\ell_0$ regularization. For each $q$ and number of dimensions, SISSO selects
the model with the smallest root mean-squared error (RMSE). We also quantify
model performance using the maximum absolute error (MaxAE) and the square of the
Pearson correlation coefficient, $r^2$.

\subsection{Feature construction and training}
We apply SISSO to three easily accessible and interpretable primary features:
the water/octanol partitioning free energy, $\dgwol$, and the acid dissociation
constants \apkatext and \bpkatext as provided by Menichetti
\etal\cite{menichetti2019drug} We thereby seek a refinement or correction to the
commonly used model based on hydrophobicity alone.\cite{Levin1980} The mean
absolute errors associated with the $\dgwol$ and \pkatext predictions (0.36
kcal/mol\cite{tetko2002application} and 0.86 units,\cite{Liao2009} respectively)
make them reliable primary descriptors. To avoid constructing features with
different units, we multiply the partitioning free energy by the inverse
temperature: $\beta\dgwol$, using $T=300$\,K following Menichetti
\etal\cite{menichetti2019drug} Starting with the set of primary features $\Phi_0
= \{\beta\dgwol, \apkamath, \bpkamath\}$, we consider the construction of
secondary features for up to two iterations (\ie $q=2$), where $\Phi_1$ and
$\Phi_2$ consist of roughly $30$ and $2\,000$ features, respectively. We limit
the SISSO screening size to $500$ and consider up to three-dimensional
descriptors. We train on 10\% of the available data (see SI for the input
script), and use the remaining 90\% for hold-out evaluation. We draw these
train/test sets uniformly at random, without replacement. To reduce variance, we
report the average performance over ten independently drawn train/test sets.
Ouyang \etal\cite{ouyang2018sisso} emphasized that SISSO works reliably when the
number of materials in a training set, $N$, sufficiently exceeds the number of
candidate features. In particular, SISSO requires $N\geq k d \ln(\#\Phi)$, where $k \sim
1$--$10$, $d$ is the dimension, and $\#\Phi$ is the size of the feature space.
For $\Phi_2$, the relevant feature space used to train our model, the relation
requires $N \geq 10\cdot 2\cdot\ln(2\cdot 10^3) \simeq 150$. By training our
models with only 10\% of the dataset ($\sim 42\,000$ compounds), SISSO is well
within a regime to provide meaningful and consistent results.

\section{Results and Discussions}

\subsection{Learning permeability models}
Table~\ref{tab:descriptors} contains the four models considered in this work:
($i$) $f^{\rm Hyd}$ is a baseline hydrophobicity model, which linearly
correlates with water/octanol partitioning free energy; and ($ii$--$iv$) $f^{\rm
1D}$ to $f^{\rm 3D}$ linearly correlate with one to three secondary feature(s)
identified by SISSO. For each model, $c_i^{\rm m}$ correspond to non-zero
coefficients for model m and index $i$, reported in Tab.~\ref{tab:descriptors}.
For all models, $\dgwol$ takes on a central role, as expected by the performance
of the baseline. The simplest SISSO model, $f^{\rm 1D} = c_0^{\rm 1D} + c_1^{\rm
1D}(\apkamath-\bpkamath-2\beta\dgwol)$, is remarkably robust: it is
systematically identified as the best performing 1D model across all 10 training
sets. Given that we trained on only 10\% ($\sim 42\,000$ compounds) of the
dataset, this highlights this model's performance compared to all other
candidates (see SI for a list of candidate one-dimensional models). The
stability of the model---given such a small training fraction---speaks for the
robustness of the equation. Its improved accuracy compared to the baseline will
be evaluated further down.

\begin{table*}[htbp]
	\caption{Permeability models, descriptor coefficients, and model
		performance: RMSE, MaxAE (both in $\log_{10}$ units of the permeability
		coefficient), and $r^2$. The models considered are: baseline
		hydrophobicity model, $f^{\rm Hyd}$, as well as SISSO with up to $3$
		feature dimensions: $f^{\rm 1D}$, $f^{\rm 2D}$, and $f^{\rm 3D}$.
		Compared to the baseline, SISSO systematically leads to more accurate
		models. Descriptor coefficients and performance metrics are averaged
		over training and test sets, respectively. All standard errors are small
		and reported in the SI.}
	\begin{ruledtabular}
		\small
		\begin{tabular}{lll|cccc|ccc}
			\multicolumn{3}{c|}{Model} & $c_0$ & $c_1$ & $c_2$           & $c_3$
			                           & RMSE & MaxAE & $r^2$ \\
			\hline
			$f^{\rm Hyd}$              & $=c_0^{\rm Hyd}$ & $+c_1^{\rm Hyd}
				\beta\dgwol$ & $-3.444$ & $-0.648$ & & & $1.53$ & $11.82$ &
				$0.64$ \\
			$f^{\rm 1D}$               & $=c_0^{\rm 1D}$ & $+c_1^{\rm
				1D}(\apkamath-\bpkamath-2\beta\dgwol)$ & $-5.419$ & $0.163$ & &
				& $1.40$ & $6.35$ & $0.70$ \\
			$f^{\rm 2D}$               & $=c_0^{\rm 2D}$ & $+c_1^{\rm
				2D}(\sqrt[3]{\beta\dgwol}+\beta\dgwol-\apkamath)$ & $-5.753$ &
				$-0.487$ & $-0.017$                                          & &
				$1.06$ & $8.28$          & $0.83$          \\
			                           & & $+c_2^{\rm
			                           2D}(\apkamath^2+\bpkamath^2)$          &
			                           & &                 & & & & \\
			$f^{\rm 3D}$               & $=c_0^{\rm 3D}$ & $+c_1^{\rm
				3D}(\beta\dgwol-\apkamath)$ & $-7.101$ & $-0.614$ & $-0.001$ &
				$-0.018$        & $0.94$ & $8.19$ & $0.86$ \\
			                           & & $+c_2^{\rm
			                           3D}(\bpkamath^2(\apkamath+\bpkamath))$ &
			                           & & & & & & \\
			                           & & $+c_3^{\rm
			                           3D}(\apkamath^2+(\beta\dgwol)^2)$      &
			                           & &                 &                 \\
		\end{tabular}
		\label{tab:descriptors}
	\end{ruledtabular}
\end{table*}

We also report more complex 2D and 3D models in Tab.~\ref{tab:descriptors}.
While we will show below that these yield even better accuracy compared to
$f^{\rm 1D}$, they are specifically tailored to the training set used: $f^{\rm
2D}$ and $f^{\rm 3D}$ are ranked as the best model in eight and five out of the
10 training sets, meaning that other models of similar complexity closely
compete.

\subsection{Model performance}
\begin{figure}[htbp]
	\centering
	\includegraphics[width=0.9\linewidth]{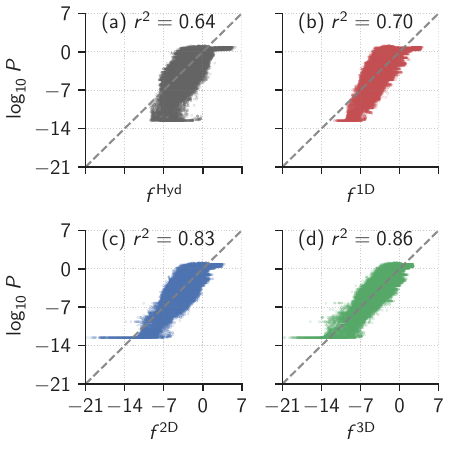}
	\caption{Performance of the four permeability models against the reference
		for the dataset of $418$k small molecules: (a) baseline hydrophobicity
		model $f^{\rm Hyd}$; (b--d) 1D to 3D data-driven SISSO models. See
		Tab.~\ref{tab:descriptors} for their expressions.}
	\label{fig:logp_sisso1d_deltaG}
\end{figure}

We now turn to the performance of these four models. Tab.~\ref{tab:descriptors}
reports their RMSE, MaxAE, and squared Pearson correlation coefficient, $r^2$,
averaged over the test sets. Going from the baseline to more complex SISSO
models, the systematic decrease in the RMSE is accompanied by an increase in the
correlation coefficient. On the other hand, the maximum absolute error does show
a clear minimum for $f^{\rm 1D}$. This offers a first hint at the appealing
balance between generalization and interpretability of the 1D SISSO model. The
performance of these four models is depicted in
Fig.~\ref{fig:logp_sisso1d_deltaG} for the entire dataset, where we report each
model against reference values. Going from baseline to SISSO models of
increasing complexity, the distribution does lean increasingly toward the $y=x$
correlation line. The presence of horizontal stripes in
Fig.~\ref{fig:logp_sisso1d_deltaG} results from the degenerate use of CG
mappings for many molecules.\cite{Kanekal2019} This artifact is most notable for
$f^{\rm Hyd}$, which solely relies on hydrophobicity, whereas the others have
chemically specific acidity information. For the 2D and 3D models, we also point
out outliers at the lowest permeability values. Fig.~S1 in the SI shows the
distribution of compounds: these low-permeability values are scarcely populated,
both in algorithmically and synthesized compound
databases.\cite{menichetti2019drug} Here they represent only $0.07\%$ of the
dataset, and our uniform sampling of training points likely brought in only few
of them. They likely result from poor extrapolation behavior of the 2D and 3D
models, which notably include powers of two of several variables.

\begin{figure*}[htbp]
	\includegraphics[width=0.9\linewidth]{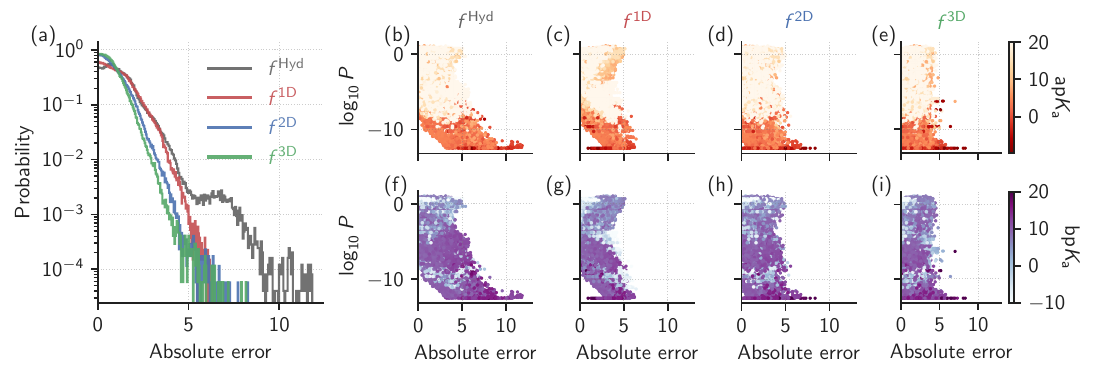}
	\caption{Absolute error analysis. (a) Error (in $\log_{10}$ units)
		distribution for all models; Error decomposed as a function of (b--e)
		\apkatext and (f--i) \bpkatext. Stronger acids/bases are shown in darker
		colors.}
	\label{fig:best_descriptors}
\end{figure*}

To better understand the performance of each model, we analyze their errors in
greater detail. Fig.~\ref{fig:best_descriptors}a shows the distribution of
absolute error. The large dataset at our disposal allows us to evaluate more
than $4$ orders of magnitude of this distribution. The comparison between
$f^{\rm Hyd}$ and $f^{\rm 1D}$ proves insightful: while they are remarkably
close up to errors of 5 $\log_{10}$ units, the baseline then displays a
significant hump, while the SISSO 1D model keeps decaying monotonously. Both
models rely on $\beta\dgwol$, which explains the remarkable agreement early on,
while the stark difference between the two curves is entirely due to the effect
of acidity. This is confirmed by a further decomposition of the error as a
function of acidity, showing that $f^{\rm Hyd}$ leads to larger errors for
stronger acids and bases (Panels b and f), while SISSO 1D significantly reduces
the error in this regime (Panels c and g).

In comparison, the more complex SISSO 2D and 3D display a shift in the error
distribution toward lower errors (Fig.~\ref{fig:best_descriptors}a) compared to
the 1D model. At low probability however, we observe a significant change in the
slope of the decay, indicating worse performance for a small number of outliers.
This is also illustrated when decomposing the error in terms of acidity in
Panels (d--i): while the overall performance improves, we identify more
outliers. These outliers mostly lie at low permeability values (reminiscent of
Fig.~\ref{fig:logp_sisso1d_deltaG}), and for strong \apkatext or \bpkatext. Poor
performance at large acidity values could take place if these were absent of the
small training fraction.

\subsection{Validation against atomistic simulations}
The SISSO models should naturally be prone to systematic errors
inherent to the training dataset. While we expect our systematic
integration of the ISDM permeability coefficient (Eq.~\ref{eq:isdm})
to ensure robust functional relationships, systematic errors in the
parameters are likely to affect the fitting coefficients. Reference
permeability values were extracted from computationally efficient
coarse-grained computer simulations, at the cost of force-field
accuracy. Still, a comparison of the coarse-grained simulations
against atomistic computer simulations had indicated an excellent
agreement of 1 $\log_{10}$ unit across a limited set of small
molecules \cite{menichetti2019drug}. Here we compare the performance
of the four permeability models against the atomistic simulations of
Carpenter \etal\cite{carpenter2014method} This set of $12$ organic
compounds covers a range of molecular weights that goes significantly
beyond our training set: an average of $243$~Da and up to $319$\,Da,
comparable to that of real drugs, given that more than 60\% of drugs
have molecular weight below 300\,Da.\cite{feher2003property} On the
other hand our training HTCG database only contained compounds up to
$160$\,Da. This thus presents a challenging test for the
generalizability of the SISSO models.

\begin{figure*}[htbp]
	\centering
	\includegraphics[width=0.9\linewidth]{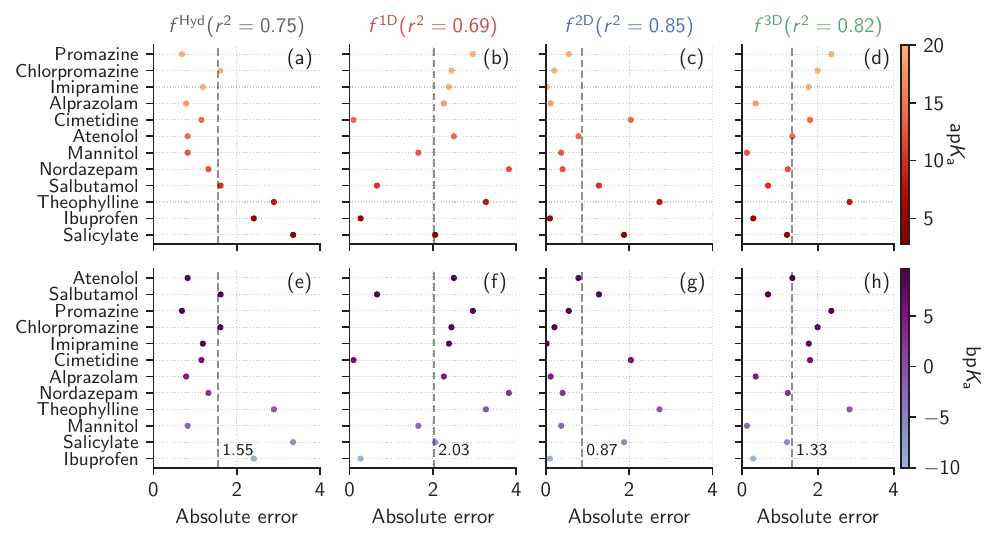}
	\caption{Absolute error (in $\log_{10}$ units) against atomistic simulations
		for $12$ reference small molecules.\cite{carpenter2014method} The error
		is displayed for all four models and as a function of (a--d) \apkatext
		and (e--h) \bpkatext.}
	\label{fig:atomistic_comparison}
\end{figure*}

Fig.~\ref{fig:atomistic_comparison} shows the absolute error against atomistic
simulations across all $12$ small molecules and for our four models. For each
model, we display the error as a function of both \apkatext and \bpkatext. The
baseline model yields absolute errors between $1$ and almost $4$ $\log_{10}$
units. While larger errors correlate with strong acids, they do not seem to
correlate with larger bases. Unlike Fig.~\ref{fig:best_descriptors}, the
minuscule set of atomistic compounds prevents us from drawing conclusions that
would hold across any significant subset of chemical space. Turning to SISSO 1D,
we observe a small but noticeable decrease in performance, where the mean
absolute error (MAE) increases from $1.55$ to $2.03$ $\log_{10}$ units. The MAE,
however, decreases against the baseline when considering the more complex SISSO
2D and 3D: $0.87$ and $1.33$~$\log_{10}$ units, respectively. The non-monotonic
decrease of the MAE with increasing complexity in
Fig.~\ref{fig:atomistic_comparison}b--d suggests the role of the small
validation dataset considered. Overall though, the incorporation of acidity does
lead to an improved reproduction of the permeability coefficient. It validates
the SISSO-derived equations on permeability coefficients derived using
independent methods and outside the chemical space of the training data.

\subsection{Acid--base asymptotes}
The analysis so far highlights how model complexity impacts accuracy. Missing
from the analysis so far is the consideration of interpretability. The two
one-dimensional models---the baseline and SISSO 1D---highlight a simple
mechanism as to the functional dependence of the permeability coefficient on
both hydrophobicity and acidity. Focusing on SISSO 1D specifically, we rewrite
the model in terms of two contributions
\begin{align}
	\label{eq:cs2}
	\notag
	f^{\rm 1D} = c_0^{\rm 1D} + c_1^{\rm 1D} \big[ & \left(\apkamath - \beta \dgwol\right)           \\
	                                               & + \left(- \bpkamath - \beta \dgwol\right)\big].
\end{align}

\begin{figure}[htbp]
	\includegraphics[width=0.9\linewidth]{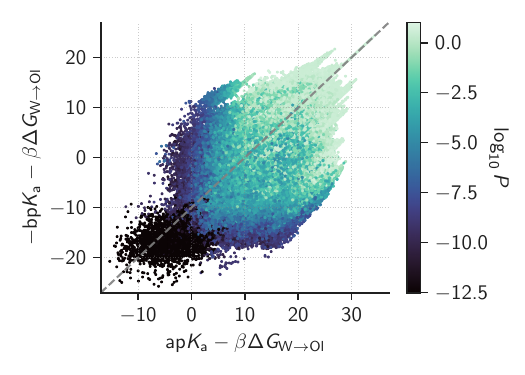}
	\caption{SISSO 1D model of passive permeability. The decomposition of the
		secondary feature in two axes highlights the role of hydrophobicity
		(along the diagonal) compared to acidity (vertical and horizontal).}
	\label{fig:sisso1d_surface}
\end{figure}

Fig.~\ref{fig:sisso1d_surface} displays the permeability coefficient as a
function of these two contributions. The symmetric contribution of $\beta
\dgwol$ in the two terms of Eq.~\ref{eq:cs2} indicates that the baseline
hydrophobicity model manifests itself along the diagonal. Notably missing from
the diagonal behavior are the dark vertical and horizontal basins. They localize
at $\apkamath - \beta \dgwol \sim 0$ and $- \bpkamath - \beta \dgwol \sim -15$,
and represent strong-acid and strong-base regimes.  In what follows, we provide
an asymptotic rationalization of the functional form of Eq.~\ref{eq:cs2}.

\begin{figure}[htbp]
	\centering
	\includegraphics[width=0.9\linewidth]{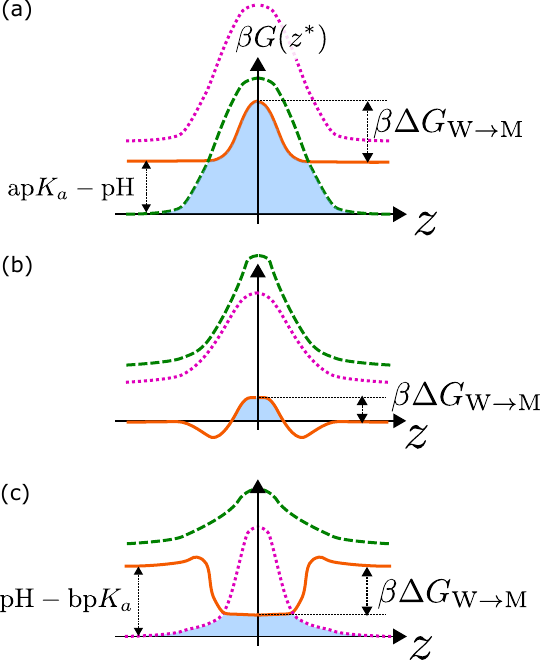}
	\caption{Sketch of the permeation mechanism in three regimes: (a) strong
		acid, (b) neutral compound, and (c) strong base. The curves display the
		neutral (solid orange), acidic (green dashed), and basic (pink dotted)
		PMFs. The blue area under the effective PMF directly links to the
		permeability coefficient.}
	\label{fig:sketch}
\end{figure}

To rationalize Eq.~\ref{eq:cs2}, we first outline the role of our three primary
descriptors in the ISDM model (Eq.~\ref{eq:isdm}). Fig.~\ref{fig:sketch}
illustrates the well known interplay between PMF and acidity, in particular how
the latter shifts the PMFs of the neutral and (de)protonated species. In the
following, we will denote the PMFs of the neutral, protonated, and deprotonated
species as $G_\mathrm{N}(z)$, $G_\mathrm{B}(z)$, and $G_\mathrm{A}(z)$,
respectively.

The difference between \apkatext or \bpkatext and \phtext dictate the propensity
for the PMFs to cross each other. The ISDM relies on a total resistivity
(defined in Eq.~\ref{eq:isdm}), $R_\textup{T}$, such that $R^{-1}_\textup{T}(z)
= R^{-1}_\mathrm{N}(z) + R^{-1}_\mathrm{B}(z) + R^{-1}_\mathrm{A}(z)$, analogous
to the total resistance in a parallel electric circuit.

The PMFs of the neutral, protonated, and deprotonated species can be linked in
water thanks to their \apkatext and \bpkatext values, as well as the \phtext of
the environment, through the following equations
\begin{align}
	\label{eq:apka}
	\beta^{-1}(\phmath - \apkamath)\ln 10 & = G_\mathrm{N}(\infty) - G_\mathrm{A}(\infty), \\
	\label{eq:bpka}
	\beta^{-1}(\phmath - \bpkamath)\ln 10 & = G_\mathrm{B}(\infty) - G_\mathrm{N}(\infty),
\end{align}
where $G(\infty)$ indicates that the compound is located in bulk water.
Eqs.~\ref{eq:apka} and \ref{eq:bpka} effectively link the difference between the
\phtext of the environment with the \pkatext of the compound to a shift in the
PMFs. Without loss of generality, we will shift all free energies such that zero
corresponds to the most favorable compound in bulk water. Eqs.~\ref{eq:apka} and
\ref{eq:bpka} allow us to explicitly link \pkatext information with the total
resistivity
\begin{align}
	R_\textup{T}^{-1}(z) = D(z) \Large[
	 & {\rm e}^{-\beta \left(G_\mathrm{N}(z) - G_\mathrm{N}(\infty)\right)} \notag   \\
	 & + {\rm e}^{-\beta \left(G_\mathrm{B}(z) - G_\mathrm{B}(\infty)\right)} \notag \\
	 & + {\rm e}^{-\beta \left(G_\mathrm{A}(z) - G_\mathrm{A}(\infty)\right)}
	\Large]\; , \label{eq:RT}
\end{align}
where we assume that all protonation states yield identical
diffusivity.\cite{menichetti2019drug} Because Eq.~\ref{eq:RT} takes on a
relatively complex form, we will consider only asymptotic regimes:
\begin{itemize}
	\item Neutral compounds entail no strong acid or base characteristic, i.e.,
	      $\apkamath  \gg  \phmath$ and $\bpkamath \ll \phmath$, such that the
	      compound is effectively unable to (de)protonate, and
	      $G_\mathrm{N}(\infty) = 0$. Eq.~\ref{eq:RT} can be simplified to
	      $R_\textup{T}(z) \approx D^{-1}(z)\exp\left[ \beta G_\mathrm{N}(z)
	      \right]$.\cite{menichetti2019revisiting}
	\item Strong acids consist of solutes that display at least one functional
	      group for which $\apkamath \ll \phmath$. For neutral \phtext ($\approx
	      7$), this would correspond to $\apkamath < 4$. We set
	      $G_\mathrm{A}(z\to\infty) = 0$. In this regime the third exponential
	      in Eq.~\ref{eq:RT} would dominate the other two, leading to $
	      R_\textup{T}(z) \approx D^{-1}(z)\exp\left[ \beta G_\mathrm{N}(z) +
	      (\phmath - \apkamath)\ln 10 \right]$.
	\item Strong bases would display at least one functional group where
	      $\bpkamath \gg \phmath$. For neutral \phtext, this would correspond to
	      $\bpkamath > 10$. We set $G_\mathrm{B}(z\to\infty) = 0$. Using a
	      similar argument, Eq.~\ref{eq:RT} would be dominated by the second
	      exponential, leading to $ R_\textup{T}(z) \approx
	      D^{-1}(z)\exp\left[\beta G_\mathrm{N}(z) - (\phmath - \bpkamath)\ln 10
	      \right]$.
\end{itemize}
The total resistivities still require integration over the reaction coordinate
$z$, which we simplify to the largest contribution of the
PMF.\cite{lee2016simulation} The effective resistivity model is equivalent to
choosing the lowest PMF at any value of $z$: $G_\textup{eff}(z) = \min_i
G_i(z)$, where $i$ runs over the neutral, protonated, and deprotonated species.
In addition, the dominating contribution of the effective permeability will come
from its maximum value, corresponding to a position $z^* =
\underset{z}{\arg\max}\, G_\textup{eff}(z)$. Assuming that the largest
contribution of the permeability arises from the total resistivity at $z^*$, we
obtain $P \approx R^{-1}_\textup{T}(z^*)$. This yields the following acid--base
asymptotic regimes
\begin{widetext}
	\begin{equation}
		\label{eq:permsimp}
		\log_{10} P \approx
		\begin{cases}
			\log_{10} D(z^*) - \frac\beta {\ln 10} G_\mathrm{N}(z^*)- \phmath + \apkamath  & \text{if } \apkamath \ll \phmath,                                    \\
			\log_{10} D(z^*) - \frac\beta {\ln 10} G_\mathrm{N}(z^*) + \phmath - \bpkamath & \text{if } \bpkamath \gg \phmath,                                    \\
			\log_{10} D(z^*) - \frac\beta {\ln 10} G_\mathrm{N}(z^*)                       & \text{if } \apkamath \gg \phmath \text{ and } \bpkamath \ll \phmath.
		\end{cases}
	\end{equation}
\end{widetext}

\begin{figure*}[htbp]
	\includegraphics{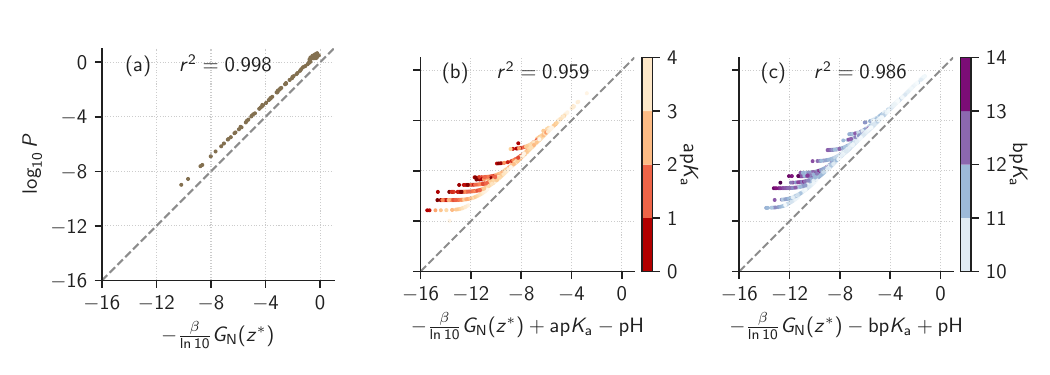}
	\caption{Simple asymptotic estimates (Eq.~\ref{eq:permsimp}) against
		reference permeability coefficients at neutral \phtext. (a) Neutral
		compounds; (b) Strong acids; and (c) Strong bases. }
	\label{fig:acid_base_neutral}
\end{figure*}

To numerically test Eq.~\ref{eq:permsimp}, we identify datapoints corresponding
to the three asymptotic regimes: the neutral compounds ($\apkamath>10$ and
$\bpkamath<4$), strong acids ($0<\apkamath<4$ and $\bpkamath <4 $), and strong
bases ($10<\bpkamath<14$ and $\apkamath >10$). For simplicity, we only
considered non-zwitterionic compounds. We assume that $\log_{10} D(z)$ yields no
significant, chemically specific effect, and uniformly shift the permeability
coefficient across the chemical space of compounds considered.
Fig.~\ref{fig:acid_base_neutral} shows the agreement between
Eq.~\ref{eq:permsimp} and the reference permeability coefficients. All show
strong linear correlation: for neutral compounds ($r^2=0.998$), strong acids
($r^2=0.959$), and strong bases ($r^2=0.986$). These results numerically
validate the asymptotes of Eq.~\ref{eq:permsimp}.

More importantly, the asymptotes provide a physically motivated rationale for
the two contributions of Eq.~\ref{eq:cs2}: $\apkamath - \beta \dgwol$ and $-
\bpkamath - \beta \dgwol$. We first note that $\dgwol$ is related to
$G_\mathrm{N}(z^*)$. The depth at which the effective PMF is the highest, $z^*$,
will almost always be close to the membrane midplane: $z^* \approx 0$. The main
exception to this are hydrophobic compounds, for which the highest point in the
PMF is in water (Fig.~\ref{fig:sketch}c). Furthermore, $G(z^*=0)$, which
corresponds to the transfer free energy from water to membrane midplane, has
been shown to linearly relate to the water/octanol partitioning free energy,
$\Delta G_{\textup{W}\to \textup{M}} \propto
\dgwol$.\cite{menichetti2017insilico} This connects $G_\mathrm{N}(z^*)$ present
in Eq.~\ref{eq:permsimp} to $\dgwol$ in Eq.~\ref{eq:cs2}. We then observe that
the asymptotes and Eq.~\ref{eq:cs2} have the same signs and exponents of
\apkatext and \bpkatext. For an acidic or a basic compound, if we consider the
compound-specific \pkatext and substitute $\dgwol$ by the compound's
$G_\mathrm{N}(z^*)$ in the relevant among the two contributions of
Eq.~\ref{eq:cs2}, we indeed recover one the asymptotes. As for neutral
compounds, $G_\mathrm{N}(z^*)$ is the only relevant quantity while estimating
permeability.

\subsection{Drug--membrane permeability across membranes}

The following analyzes how drug--membrane permeability changes according to
membrane composition. We hypothesize that the functional form of SISSO 1D is
applicable to other lipid membranes, and use it as a starting point. We take
advantage of the above-mentioned asymptotic regimes to \emph{limit} the amount
of information needed from new membranes. The regime of neutral compounds
described in Eq.~\ref{eq:permsimp} can be used advantageously because it only
requires information on neutral PMFs. We rely on the dataset of Hoffmann \etal,
which precisely contains PMF information---but no permeability---in various
membranes, and only for neutral compounds.\cite{Hoffmann2020} We specifically
analyze the change in permeability when turning to phosphocholine (PC) membranes
made of different lipids, varying in both tail length and level of unsaturation.

\begin{figure}[htbp]
	\centering
	\includegraphics[width=0.9\linewidth]{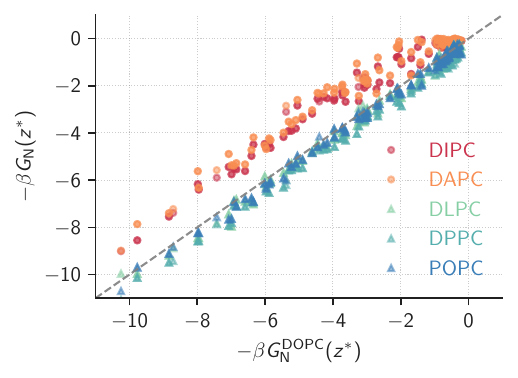}
	\caption{Variation of $-\beta G_{\rm N}(z^*)$ between the original membrane,
		DOPC, and others. Note the shift between ($i$) DLPC, DPPC, and POPC and
		($ii$) DIPC and DAPC.}
	\label{fig:correl_g_star}
\end{figure}

Fig.~\ref{fig:correl_g_star} shows the relation of $-\beta G_{\rm N}(z^*)$---the
dominant term for the permeability of neutral compounds
(Eq.~\ref{eq:permsimp})---between the original membrane used in this work, DOPC,
and others. All curves follow a line, indicating that the asymptotic regime for
neutral compounds of Eq.~\ref{eq:permsimp} holds for all membranes. We find two
families of lines with different intercepts: DLPC, DPPC, and POPC show an
intercept with DOPC that is roughly 0, while DIPC and DAPC have an intercept
that is approximately 1.4.

To better understand these results, we first recall that $G_{\rm N}(z^*)$
corresponds to the highest value of the neutral PMF. We can safely ignore
contributions of the charged PMFs, such that $G_{\rm N}(z^*)$ denotes the
highest point of the \emph{effective} PMF. The excellent agreement between DLPC
and DPPC indicates that tail length ($3$ and $4$ beads long, respectively) does
not impact the permeability. This is expected, given that tail length is only
expected to change the length, but not the height, of the hydrophobic plateau.
On the other hand, the further agreement between them and POPC and DOPC indicate
a lack of dependence on tail saturation for these lipids ($1$ and $2$
unsaturated beads, respectively). Remarkably, the shift in the intercept appears
only for DIPC and DAPC---lipids that exhibit more unsaturations: $4$ and $8$,
respectively. Notably, they display the \emph{same} shift in $-\beta G_{\rm
N}(z^*)$. As such, the level of unsaturation is a determining factor, but does
not gradually impact $-\beta G_{\rm N}(z^*)$.

Interestingly, previous Martini studies of ternary membranes with DPPC and
cholesterol have shown that DIPC and DAPC strongly phase separate into a
liquid-disordered (Ld) phase.\cite{Rosetti2012, Davis2013, Pantelopulos2018}
Inconsistent conclusions were drawn from different studies with DOPC, pointing
to a thermodynamic drive that is weak, at best.\cite{baoukina2012molecular,
Davis2013, girard2020regulating} As for POPC, there is no sign of phase
transition.\cite{Davis2013} The results on Fig.~\ref{fig:correl_g_star} mirror
these trends: we find a clear shift of $-\beta G_{\rm N}(z^*)$ depending on the
ability of the membrane to form an Ld domain. The Ld-domain formation of course
hinges on the presence of DPPC and cholesterol, which are notably absent from
our reference simulations.\cite{Hoffmann2020} The trends are surprising in that
they show a dependence on lipid-tail unsaturation that is \emph{stepwise} rather
than proportional.  While we defer a more detailed study to future work, we
suggest the role of the entropic character of the lipid-tail fluctuations.

Other studies before us have reported a clear change in permeability between Ld
and Lo domains: Ghysels \etal used both atomistic simulations and electron
paramagnetic resonance spectroscopy experiments on the permeation of water and
oxygen, and found a permeability ratio  $P({\rm Ld})/P({\rm Lo}) \approx
3$.\cite{Ghysels2019} Here the CG simulations yield a shift in $-\beta G_{\rm
N}(z^*)$, which translates to a ratio of the permeability coefficients of
$\approx 25$. Our estimates are thus within one $\log_{10}$ unit of the results
of Ghysels \etal for their specific compounds. The mechanism remains to be
clearly identified, although this could be consistent with the proposed role of
local membrane surface density (i.e., its propensity to form transient
holes).\cite{Liu2019}

To summarize, our use of single-component lipid membranes only allows us to
speculate as to the shift in Fig.~\ref{fig:correl_g_star}. The link to Lo/Ld
domain formation is in line with prior atomistic simulations and experiments for
specific compounds.\cite{Ghysels2019} In the broader context, our results could
help generalize their results across the chemical space of drugs. The results
also suggest simple \emph{additive} corrections to our effective equation when
considering different membranes.

\section{Conclusions}
We propose to learn a functional relationship between hydrophobicity and acidity
as a simple surrogate for passive membrane permeability. Our approach, combining
symbolic regression and compressed sensing, is data-driven \emph{and}
interpretable, and based on large databases of high-throughput coarse-grained
simulations. Sure-Independence Screening and Sparsifying Operator (SISSO) builds
a hierarchy of models of increasing complexity. Models prove increasingly
accurate, yet more complex models are more prone to discrepancies for a few
outliers. Our SISSO 1D model offers improved accuracy compared to the
hydrophobicity baseline, and yet excellent interpretability. We identify the
simple and interpretable equation $f^{\rm 1D} = c_0^{\rm 1D} + c_1^{\rm
1D}(\apkamath-\bpkamath-2\beta\dgwol)$, where \apkatext and \bpkatext
characterize acidity, $\dgwol$ is the water/octanol partitioning coefficient,
and $c_0^{\rm 1D}$ and $c_1^{\rm 1D}$ are the only two fitting parameters. We
rationalize the model by an analysis of the asymptotic regimes of the
inhomogeneous solubility--diffusion model (ISDM). The asymptotes are validated
numerically and confirm the SISSO 1D equation, implicitly testifying to the
accuracy of the underlying HTCG resolution. Broad agreement numerically
validates the use of a single bulk hydrophobicity measure to effectively replace
the potential of mean force, which has been exploited by others
before.\cite{lee2016simulation} Importantly, the interplay of hydrophobicity
together with acidity leads to a significant improvement in model accuracy of
the SISSO 1D equation for ionizable groups. The SISSO equations show
improvements over a challenging set of compounds that are much larger than the
training set. Critically, our work refines the common role of hydrophobicity in
passive permeation to relate it \emph{functionally} with acidity.

The separation of the ISDM in asymptotic regimes allows us to build
drug--permeability models across membranes with limited information
only. Using only the potential of mean force of neutral solutes, we
infer the change in permeability for membranes with lipids of varying
tail length and level of unsaturation. We observe a surprising change
in permeability coefficient: lipid-tail unsaturation contributes
stepwise rather than proportionally. Our findings are in line with
recent atomistic simulations and electron paramagnetic resonance
spectroscopy experiments, highlighting the distinction between lipids
primarily involved in liquid-disordered (Ld) and liquid-ordered (Lo)
domains. The approach offers a data-driven, interpretable analysis of
drug--membrane passive permeability across both drugs and membranes.

\section*{Supplementary material}
See supplementary material for definitions of dissociation constants (\apkatext
and \bpkatext), distribution of compounds across permeability,
Tab.~\ref{tab:descriptors} along with the error values, list of best
one-dimensional descriptors, and the SISSO input script.

\section*{Data availability}
In this study, we used the database provided by Menichetti \etal
\cite{menichetti2019drug} It is openly available at
\url{https://doi.org/10.1021/acscentsci.8b00718}.

\begin{acknowledgments}
	The authors thank Oleksandra Kukharenko, Roberto Menichetti, and Yasemin
	Bozkurt Varolg{\"u}ne{\c{s}} for critical reading of the manuscript and
	Kiran Kanekal and Martin Girard for insightful discussions. Data analysis
	relied extensively on the open source packages Numpy,\cite{harris2020array}
	Matplotlib,\cite{hunter2007matplotlib} and Pandas.\cite{reback2020pandas,
	mckinney2010data} We acknowledge support by BiGmax, the Max Planck Society's
	Research Network on Big‐Data‐Driven Materials‐Science. T.B. was partially
	supported by the Emmy Noether program of the Deutsche Forschungsgemeinschaft
	(DFG).
\end{acknowledgments}

\bibliography{paper}
\end{document}


\title{Supplementary information for ``Data-driven equation for drug--membrane permeability across drugs and membranes''}

\author{Arghya Dutta}
\email{dutta@mpip-mainz.mpg.de}
\affiliation{Max Planck Institute for Polymer Research, Mainz, Germany}

\author{Jilles Vreeken}
\affiliation{CISPA Helmholtz Center for Information Security, Saarbr\"ucken, Germany}
\author{Luca Ghiringhelli}
\affiliation{The NOMAD Laboratory at the Fritz Haber Institute of the Max Planck
Society and Humboldt University, Berlin, Germany}

\author{Tristan Bereau}
\affiliation{Max Planck Institute for Polymer Research, Mainz, Germany}
\affiliation{Van 't Hoff Institute for Molecular Sciences and Informatics
    Institute, University of Amsterdam, Amsterdam, The Netherlands}

\date{\today}
\maketitle
\section{Definition of the ionization constants and \pkatext}
Menichetti~\emph{et. al.}\cite{menichetti2019drug} followed the convention of
ChemAxon \cite{chemaxonurl} while defining \apkatext and \bpkatext. As it is a
bit different from the usual of acidic and basic $\pkamath$s, here we provide
the details. The ionization constant \katext and \pkatext are defined as
\begin{align}
    \ce{K_a  & = \frac{[conjugate base]\times[H+]}{[conjugate acid]}},                     \\
    \ce{pK_a & = -\log_{10}K_a = pH + \log_{10}\frac{[conjugate acid]}{[conjugate base]}},
\end{align}
where \ce{pH = -\log_{10}[H+]}. In their simulations, Menichetti~\emph{et. al.}
always started from a neutral compound which, depending on the pH and \pkatext,
can either protonate or deprotonate. We consider these two cases separately as
follows.

\subsection{Deprotonation}
A charge-neutral acid \ce{AH} can deprotonate to release a proton and a charged
conjugate base \ce{A-} by the following reaction
\begin{center}
    \ce{AH <=> A- + H+}.
\end{center}
The corresponding acidic \pkatext, denoted as \apkatext, is defined as
\begin{align}
    \ce{K_a   & = \frac{[A^-][H+]}{[AH]}},                                     \\
    \apkamath & = -\log_{10} \ce{K_a} = \ce{pH + \log_{10}\frac{[AH]}{[A^-]}}.
    \label{eq:apka}
\end{align}
\subsection{Protonation}
A charge-neutral base \ce{B} can protonate and becomes a charged conjugate acid
\ce{[BH+]} by the following reaction
\begin{center}
    \ce{BH+ <=> B + H+}.
\end{center}
To comply with the unified definition of \pkatext of ChemAxon---it is the ratio
of conjugate acid to conjugate base---the corresponding basic \pkatext, denoted
as \bpkatext, is defined as
\begin{align}
    \ce{K_a   & = \frac{[B][H+]}{[BH+]}}                                     \\
    \bpkamath & = -\log_{10} \ce{K_a} = \ce{pH + \log_{10}\frac{[BH+]}{[B]}}
    \label{eq:bpka}
\end{align}
The usefulness of this definition is that now both \apkatext and \bpkatext are
written as \ce{pK_a = pH + \log_{10}\frac{[conjugate acid]}{[conjugate
                base]}}.

Strong acids, as defined in the Acid--base asymptotes section of the paper, have
low $(\apkamath \leq 4)$. At $\phmath=7$, from Eq.~\ref{eq:apka} we find that
\begin{align}
    \ce{\frac{[AH]}{[A^-]}} = 10^{\apkamath-7} \leq 10^{-3}.
\end{align}
So, with \emph{decreasing} \apkatext, the acid's concentration \ce{[AH]} will
keep decreasing and the conjugate base's concentration \ce{[A-]} will keep
increasing, as expected. Conversely, strong bases have high $(\bpkamath \geq
    10)$. At $\phmath=7$, from Eq.~\ref{eq:bpka} we get
\begin{align}
    \ce{\frac{[BH+]}{[B]}} = 10^{\bpkamath-7} \geq 10^{3}.
\end{align}
So, with \emph{increasing} \bpkatext, the base's concentration \ce{[B]} will
keep decreasing and the conjugate acid's concentration \ce{[BH+]} will keep
increasing, again as expected.

\section{Distribution of compounds across permeability}
\begin{figure*}[!htbp]
    \includegraphics[width=0.5\linewidth]{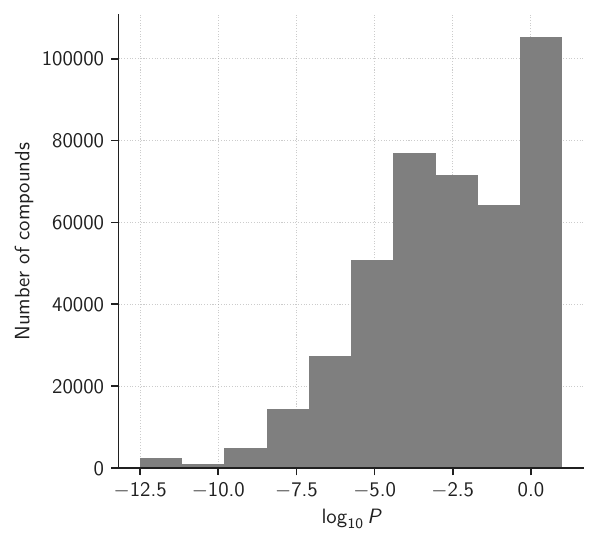}
    \caption{Distribution of the small molecules considered in this work across the range of permeability values.}
\end{figure*}
\newpage
\section{Table 1 with error values}

\begin{table*}[!htbp]
    \caption{Table I from the main text along with the standard errors shown in parentheses.}
    \begin{ruledtabular}
        \small
        \begin{tabular}{lll|cccc|ccc}
            \multicolumn{3}{c|}{Model}             & $c_0$            & $c_1$
                                                   & $c_2$            & $c_3$           & RMSE
                                                   & MaxAE
                                                   & $r^2$                                             \\
            \hline
            $f^{\rm Hyd}$                          & $=c_0^{\rm Hyd}$ & $+c_1^{\rm Hyd}
                \beta\dgwol$
                                                   & $-3.444$         & $-0.648$        &
                                                   &
                                                   & $1.53$           & $11.82$         & $0.64$       \\
                                                   &                  &
                                                   & $(\pm0.004)$     & $(\pm0.001)$    &
                                                   &                  & $(\pm 0.00)$    & $(\pm 0.00)$
                                                   & $(\pm 0.00)$                                      \\
            $f^{\rm 1D}$                           & $=c_0^{\rm 1D}$  & $+c_1^{\rm
                1D}(\apkamath-\bpkamath-2\beta\dgwol)$
                                                   & $-5.419$         & $0.163$         &
                                                   &                  & $1.40$
                                                   & $6.35$           & $0.70$                         \\
                                                   &                  &
                                                   & $(\pm0.007)$     & $(\pm0.000)$    &
                                                   &
                                                   & $(\pm 0.00)$     & $(\pm 0.02)$    &
            $(\pm0.00)$                                                                                \\
            $f^{\rm 2D}$                           & $=c_0^{\rm 2D}$  & $+c_1^{\rm
                2D}(\sqrt[3]{\beta\dgwol}+\beta\dgwol-\apkamath)$
                                                   & $-5.753$         & $-0.487$        & $-0.017$
                                                   &
                                                   & $1.06$           & $8.28$          & $0.83$       \\
                                                   &                  & $+c_2^{\rm
            2D}(\apkamath^2+\bpkamath^2)$          &
            $(\pm0.007)$                           & $(\pm0.001)$
                                                   & $(\pm0.000)$     &                 &
            $(\pm0.00)$                            & $(\pm0.03)$
                                                   & $(\pm0.00)$
            \\
            $f^{\rm 3D}$                           & $=c_0^{\rm 3D}$  & $+c_1^{\rm
            3D}(\beta\dgwol-\apkamath)$            & $-7.101$         &
            $-0.614$                               & $-0.001$         & $-0.018$
                                                   & $0.94$           & $8.19$          & $0.86$       \\
                                                   &                  & $+c_2^{\rm
            3D}(\bpkamath^2(\apkamath+\bpkamath))$ &
            $(\pm0.007)$                           & $(\pm0.002)$
                                                   & $(\pm0.000)$     & $(\pm0.000)$    &
            $(\pm0.00)$                            & $(\pm0.02)$
                                                   & $(\pm0.00)$
            \\
                                                   &                  & $+c_3^{\rm
                3D}(\apkamath^2+(\beta\dgwol)^2)$
                                                   &                  &                 &
                                                   &                                                   \\
        \end{tabular}
        \label{tab:descriptors}
    \end{ruledtabular}
\end{table*}

\section{One-dimensional descriptors}

\begin{table*}[!htbp]
    \caption{Best one-dimensional descriptors for ten training sets as predicted
        by SISSO. Each column corresponds to a particular training set. The data
        demonstrates the robustness of the predictions across training
        sets---only twelve unique descriptors are present the best ten
        descriptors from all training sets. The top three descriptors do not
        change. The best one-dimensional descriptor
        $((\apkamath-\beta\dgwol)-(\bpkamath+\beta\dgwol))$ and the baseline
        hydrophobicity descriptor have been highlighted for reference.}
    \begin{ruledtabular}
        \begin{tabular}{c|cccccccccc}
            \multirow{2}{*}{1D descriptor}                                  &
            \multicolumn{10}{c}{Rank in training set number}
            \\
                                                                            & 1
                                                                            & 2
                                                                            & 3
                                                                            & 4
                                                                            & 5
                                                                            & 6
                                                                            & 7
                                                                            & 8
                                                                            & 9
                                                                            & 10
            \\\hline
            $\highlight{((\apkamath-\beta\dgwol)-(\bpkamath+\beta\dgwol))}$ & 1
                                                                            & 1
                                                                            & 1
                                                                            & 1
                                                                            & 1
                                                                            & 1
                                                                            & 1
                                                                            & 1
                                                                            & 1
                                                                            & 1
            \\
            $((\bpkamath)^2+(\apkamath\cdot\beta\dgwol))$                   & 2
                                                                            & 2
                                                                            & 2
                                                                            & 2
                                                                            & 2
                                                                            & 2
                                                                            & 2
                                                                            & 2
                                                                            & 2
                                                                            & 2
            \\
            $(\sqrt[3]{(\bpkamath)}+\beta\dgwol)$                           & 3
                                                                            & 3
                                                                            & 3
                                                                            & 3
                                                                            & 3
                                                                            & 3
                                                                            & 3
                                                                            & 3
                                                                            & 3
                                                                            & 3
            \\
            $((\apkamath-\beta\dgwol)-\bpkamath)$                           & 4
                                                                            & 5
                                                                            & 4
                                                                            & 6
                                                                            & 4
                                                                            & 4
                                                                            & 5
                                                                            & 6
                                                                            & 5
                                                                            & 6
            \\
            $(\beta\dgwol+(\bpkamath+\beta\dgwol))$                         & 5
                                                                            & 4
                                                                            & 6
                                                                            & 7
                                                                            & 5
                                                                            & 5
                                                                            & 4
                                                                            & 4
                                                                            & 4
                                                                            & 7
            \\
            $(\sqrt[3]{(\apkamath)}-\beta\dgwol)$                           & 6
                                                                            & 7
                                                                            & 7
                                                                            & 5
                                                                            & 7
                                                                            & 7
                                                                            & 7
                                                                            & 7
                                                                            & 7
                                                                            & 5
            \\
            $(\beta\dgwol+\sqrt[3]{(\beta\dgwol)})$                         & 7
                                                                            & 6
                                                                            & 5
                                                                            & 4
                                                                            & 6
                                                                            & 6
                                                                            & 6
                                                                            & 5
                                                                            & 6
                                                                            & 4
            \\
            $(\sqrt[3]{(\beta\dgwol)}-\sqrt[3]{(\apkamath)})$               & 8
                                                                            & 8
                                                                            & 8
                                                                            & 8
                                                                            & 8
                                                                            & 8
                                                                            & 8
                                                                            & 8
                                                                            & 8
                                                                            & 8
            \\
            $((\apkamath)^{-1}+\beta\dgwol)$                                & 9
                                                                            & -
                                                                            & 10
                                                                            & -
                                                                            & 10
                                                                            & -
                                                                            & -
                                                                            & -
                                                                            & -
                                                                            & -
            \\
            $\highlight{(\beta\dgwol)}$                                     & 10
                                                                            & 9
                                                                            & 9
                                                                            & 9
                                                                            & 9
                                                                            & 9
                                                                            & 9
                                                                            & 9
                                                                            & 9
                                                                            & 9
            \\
            $(\sqrt[3]{(\beta\dgwol)}+(\bpkamath+\beta\dgwol))$             & -
                                                                            & 10
                                                                            & -
                                                                            & -
                                                                            & -
                                                                            & 10
                                                                            & -
                                                                            & 10
                                                                            & 10
                                                                            & 10
            \\
            $((\apkamath)^{-1}-\beta\dgwol)$                                & -
                                                                            & -
                                                                            & -
                                                                            & 10
                                                                            & -
                                                                            & -
                                                                            & 10
                                                                            & -
                                                                            & -
                                                                            & -
        \end{tabular}
    \end{ruledtabular}
\end{table*}

\newpage
\section{Input script}
To learn the permeability equations, we used the following
SISSO\cite{ouyang2017sisso} input script. It trains on 10\% ($=41\,897$) of the
compounds that map to a two-bead Martini representation (there are $418\,971$
such compounds) in the data provided by Menichetti
\etal\cite{menichetti2019drug}.

\begin{verbatim}
!>>>>>>>>>>>>>>>>>>>>>>>>>>>>>>>>>>>>>>>>>>>>>>>>>>>>>>>>>>>>>>>>>>>>>>>>>>>>>>>>>>>>>>>>>>>>>>>>>>
! keywords for the target properties
!>>>>>>>>>>>>>>>>>>>>>>>>>>>>>>>>>>>>>>>>>>>>>>>>>>>>>>>>>>>>>>>>>>>>>>>>>>>>>>>>>>>>>>>>>>>>>>>>>>
ptype=1  ! property type 1: continuous for regression, 2: categorical for classification
ntask=1  ! number of tasks (properties or maps) 1: single-task learning, >1: multi-task learning
nsample=41897  ! number of samples for each task (separate the numbers by comma for ntask >1)
task_weighting=1  ! 1: no weighting (tasks treated equally) 2: weighted by #sample_task_i/total_sample
desc_dim=3  ! dimension of the descriptor (<=3 for classification)
restart=.false.  ! set .true. to continue a job that was stopped but not yet finished

!>>>>>>>>>>>>>>>>>>>>>>>>>>>>>>>>>>>>>>>>>>>>>>>>>>>>>>>>>>>>>>>>>>>>>>>>>>>>>>>>>>>>>>>>>>>>>>>>>>
! keywords for feature construction and sure independence screening
! implemented operators:(+)(-)(*)(/)(exp)(exp-)(^-1)(^2)(^3)(sqrt)(cbrt)(log)(|-|)(scd)(^6)(sin)(cos)
! scd: standard Cauchy distribution
!>>>>>>>>>>>>>>>>>>>>>>>>>>>>>>>>>>>>>>>>>>>>>>>>>>>>>>>>>>>>>>>>>>>>>>>>>>>>>>>>>>>>>>>>>>>>>>>>>>
nsf=3   ! number of scalar features (one feature is one number for each material)
rung=2  ! rung (<=3) of the feature space to be constructed (times of applying the opset recursively)
opset='(+)(-)(*)(/)(exp)(log)(^-1)(^2)(^3)(sqrt)(cbrt)'  ! ONE operator set for feature transformation
maxcomplexity=10  ! max feature complexity (number of operators in a feature)
dimclass=  ! group features according to their dimension/unit; those not in any () are dimensionless
maxfval_lb=1e-3  ! features having the max. abs. data value <maxfval_lb will not be selected
maxfval_ub=1e5  ! features having the max. abs. data value >maxfval_ub will not be selected
subs_sis=500  ! size of the SIS-selected (single) subspace for each descriptor dimension

!>>>>>>>>>>>>>>>>>>>>>>>>>>>>>>>>>>>>>>>>>>>>>>>>>>>>>>>>>>>>>>>>>>>>>>>>>>>>>>>>>>>>>>>>>>>>>>>>>
! keywords for descriptor identification via a sparsifying operator
!>>>>>>>>>>>>>>>>>>>>>>>>>>>>>>>>>>>>>>>>>>>>>>>>>>>>>>>>>>>>>>>>>>>>>>>>>>>>>>>>>>>>>>>>>>>>>>>>>
method='L0'  ! sparsification operator: 'L1L0' or 'L0'; L0 is recommended!
L1L0_size4L0= 1  ! If method='L1L0', specify the number of features to be screened by L1 for L0
fit_intercept=.true.  ! fit to a nonzero intercept (.true.) or force the intercept to zero (.false.)
metric='RMSE'  ! for regression only, the metric for model selection: RMSE,MaxAE
nm_output=100  ! number of the best models to output
\end{verbatim}

\bibliography{si}